# $CuIn(Se,Te)_2$ absorbers with bandgaps < 1 eV for bottom cells in tandem applications

Thomas Paul Weiss[1], Mohit Sood[1], Aline Vanderhaegen[1], Susanne Siebentritt[1]

[1] Laboratory for Photovoltaics, Department of Physics and Materials Science, University of Luxembourg, 41, rue du Brill, L-4422 Belvaux, Luxembourg

**Abstract**

**Thin-film solar cells reach high efficiencies and have a low carbon footprint in production. Tandem solar cells have the potential to significantly increase the efficiency of this technology, where the bottom-cell is generally composed of a $Cu(In,Ga)Se_2$ absorber layer with bandgaps around 1 eV or higher. Here, we investigate $CuIn(Se_{1-x}Te_x)_2$ absorber layers and solar cells with bandgaps below 1 eV, which will bring the benefit of an additional degree of freedom for designing current-matched 2-terminal tandem devices. We report that $CuIn(Se_{1-x}Te_x)_2$ thin films can be grown single phase by co-evaporation and that the bandgap can be reduced to the optimum range (0.92 – 0.95 eV) for a bottom cell. From photoluminescence spectroscopy it is found that no additional non-radiative losses are introduced to the absorber. However, $V_{OC}$ losses occur in the final solar cell due to non-optimised interfaces. Nevertheless, a record device with 9 % power conversion efficiency is demonstrated with a bandgap of 0.96 eV and $x = 0.15$. Interface recombination is identified as a major recombination channel for larger Te contents. Thus, further efficiency improvements are anticipated for improved absorber/buffer interfaces.**

## 1. Introduction

Compound thin film photovoltaics (PV) enable high power conversion efficiency in combination with a low carbon footprint [1] and are therefore an important technology to combat the climate crisis. Ideal single-junction solar cells are limited to power conversion efficiencies around 33 % set by the Shockley-Queisser limit [2], whereas multi-junction solar cells allow to reach higher efficiencies. The reasons are lower thermalization losses and the possibility to use a larger range of the solar photon spectrum. The highest efficiencies are indeed achieved by multi-junction solar cells [3]. With the success of thin-film perovskite solar cells, these absorbers are used as top cells and enabled efficient tandem solar cells with $Cu(In,Ga)Se_2$ [4,5] or Si as a bottom cell [3,6].

An advantage of compound thin films is the tunability of the bandgap. For instance, the bandgap of the chalcopyrite family $ABC_2$ can be tuned by alloying on the different atomic sites, for instance by $(Cu,Ag)(In,Ga)(S,Se,Te)_2$, where A, B and C denote a group I, group III and group IV atom. Therefore, it is possible to adapt the bottom cell's bandgap according to the bandgap of the perovskite top cell with highest efficiency and stability. Figure 1 shows a calculation of the theoretical efficiencies for an idealized tandem solar cell either in 4-terminal (4T) or 2-terminal (2T) configuration as a function of the top- and bottom-cell bandgaps (see Appendix A for details). Both architectures allow efficiencies above 45 % in the ideal case. The region of high efficiency is larger for the 4T configuration as no current-matching constraints apply. However, the necessity of transparent conductive oxides for the back side of the top cell and the front size of the bottom imposes optical and resistive losses for such a structure, in particular for modules. These losses can be avoided for the 2T architecture. The bottom part in Figure 1 shows the maximum possible efficiency as a function of the bottom cell band gap for fixed top cell bandgaps. For the 2T architecture and already indicated by the efficiency map, it is seen that the efficiency drops quickly for bottom-cell bandgaps away from the optimum. For a top-cell bandgap of 1.6 eV, where for instance good and stable perovskite solar cells can be fabricated [7], the bottom cell's bandgap should be between 0.92 and 0.95 eV. Alloying Ag, Ga, or S to $CuInSe_2$ will only increase the absorbers bandgap above 1.0 eV. However, the $CuIn(Se_{1-x}Te_x)_2$ has a strong bandgap bowing resulting in smallest bandgaps at x = 0.5 [8]. Indeed, it is reported that $CuIn(Se_{1-x}Te_x)_2$ can reach bandgap values as low as 0.86 [9] – 0.88 [10] eV for x = 0.5 and is thus a suitable candidate for a bottom-cell in a 2T tandem architecture. Also, having a top-cell bandgap of 1.5 eV (orange line in Figure 1), the second maximum for the 2T configuration still allows efficiencies of approximately 40 %. Noteworthy, also other thin-film material systems are a good match to bottom-cell bandgaps in this energy range. In Ref. [11] a $Cu(In,Ga)S_2$ devices with a bandgap of 1.55 eV is demonstrated reaching a power conversion efficiencies of 15.5 %. CdTe reaches efficiencies of 22.3 % [12], whereas it is noted that these absorber layers are alloyed with Se, which reduces the bandgap outside the optimum range (for bottom-cell bandgaps possible with $CuIn(Se_{1-x}Te_x)_2$) to 1.45 eV.

Noteworthy, the efficiencies calculated in Figure 1 are based on complete absorption for photons with energies larger than the respective bandgaps. To obtain current-matching by adapting the top- and bottom-cell bandgaps is also the preferred solution due to a large annual energy yield [13]. In contrast, one of the record $Cu(In,Ga)Se_2$/perovskite tandem uses a perovskite top-cell absorber layer with reduced thickness to achieve current-matching [14]. Thus, the device would benefit from a bottom-cell with lowered bandgap to achieve current matching in terms of the annual energy yield.

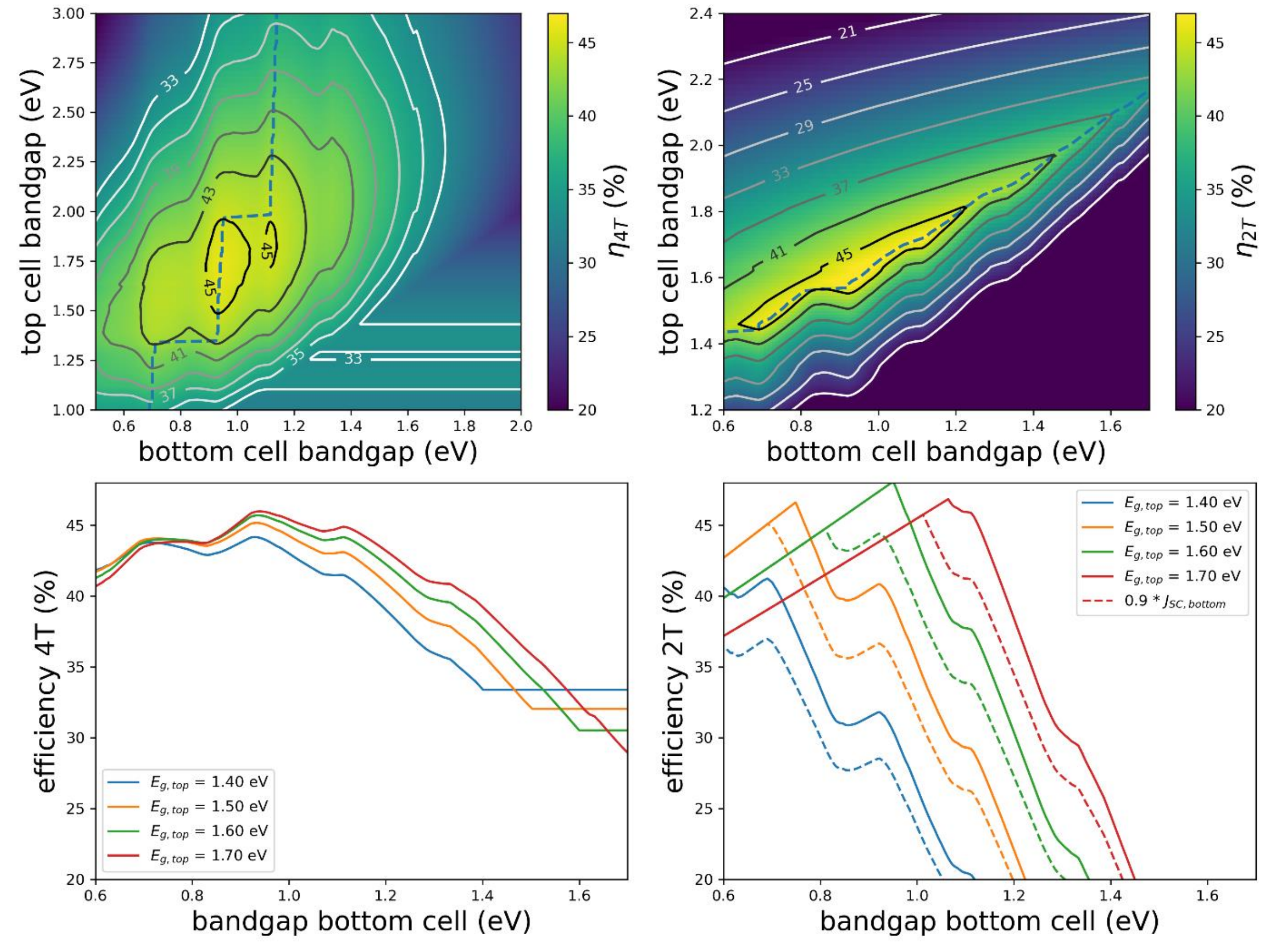

Figure 1 – calculated tandem efficiencies for a 4 terminal (left) and a 2 terminal (right) configuration as a function of the bandgaps for the top and bottom cell (top). The dashed line in the top maps indicates the optimal choice of the bottom cell bandgap to reach the highest efficiency as a function of the top cell bandgap (or vice versa).

The mixing enthalpy of the $CuIn(Se_{1-x}Te_x)_2$alloy is calculated to be $\Delta H(Se, Te) = 13$ meV/atom and thus indicates phase separation at T = 0 K [8]. However, the phase is stabilized by entropy at finite temperatures, which suggests that the mixed chalcopyrite phase is miscible in the whole compositional range [8]. Experimentally, Avon et al. used sealed quartz capsules and annealed to equilibrium at 600 – 800 °C various compositions belonging to the chalcopyrite family $(Cu,Ag)(In,Ga)(Se,Te)_2$ [15]. It is found that $CuIn(Se_{1-x}Te_x)_2$ is miscible in the complete compositional range. Li-Kao et al. used elemental co-evaporation to grow thin films of $CuIn(Se_{1-x}Te_x)_2$ employing either a single-stage or 3-stage process [10,16]. In the single-stage process all elements are evaporated with constant fluxes. In contrast, in the 3-stage process, the absorber goes from Cu-poor to Cu-rich and subsequently again to Cu-poor composition [17]. To achieve a single-phase absorber layer, it is found that single-stage absorbers can only be grown up to x=0.2, while 3-stage grown absorbers could accommodate a Te content up to x = 0.4. Finally, a solar cell is demonstrated with a bandgap of 0.89 eV with x = 0.4 and a power conversion efficiency of 7.0 % [16].

Here, we present a detailed study of loss mechanisms of $CuIn(Se_{1-x}Te_x)_2$ (compared to $CuInSe_2$) absorber layers and solar cells. We employ a single-stage co-evaporation process and grow absorbers with x up to 0.5. The quality of these absorber layers is characterized by photoluminescence spectroscopy as a function of the Te content. From these absorber layers, solar cells are fabricated and further losses are characterized, which occur upon finishing the absorber layers into solar cells. A record solar cell is demonstrated with x = 0.15, a bandgap of 0.96 eV (close to the optimum, see Figure 1) and a power conversion efficiency of 9.0%. Finally, performance bottlenecks are pin pointed, which give suggestions for future improvements of these absorber layers and solar cells.

## 2. Methods

$CuIn(Se_{1-x}Te_x)_2$ absorber layers are grown by co-evaporation of Cu, In, Se and Te from elemental sources. Elemental fluxes are changed by adjusting the respective source temperature, which is then kept constant during the deposition, i.e. the absorber layers are grown in a 1-stage process. The $CuInSe_2$ absorber layer, which will serve as a reference in this study, is grown using a 3-stage process [17]: In the first stage, only In and Se are evaporated. In the second stage, only Cu and Se are evaporated bringing the absorber to a Cu-rich composition, i.e. [Cu]/[In] > 1. In the third stage, again only In and Se are evaporated and a Cu-poor integral composition is targeted. Within each deposition process, absorber layers are grown on soda lime glass (SLG) substrates as well as SLG/Mo substrates. Substrates are heated radiatively by lamps from the back side. The substrate temperature is set between 550 °C and 570 °C for the samples grown here as measured by thermocouples in close proximity to the backside of the substrates. The deposition time is set between 60 to 80 minutes. After the deposition of the absorber layers, the samples are cooled down naturally with a maximum cooling rate of 30 °C/minute. During cooldown, Te and Se evaporation is continued. Evaporation of Te is stopped at a substrate temperature of 400 °C. At 250 °C the samples are taken out of the evaporation chamber into the loadlock, which effectively stopped the Se evaporation at 250 °C.

Composition is measured by energy dispersive X-ray analysis in an electron microscope using elemental standards. The acceleration voltage is set to 20 kV. Generally, samples with a transparent SLG substrates have a higher Te content. The reason might be a lower substrate temperature due to lower absorption of the radiation from the substrate heat source.

Structural analysis is carried out by X-ray diffraction (XRD) in Bragg-Brentano configuration with 2Θ ranging from 10 ° to 90 ° with a step-size of 0.02 °.

An approximately 40 nm thick CdS buffer layer is grown by chemical bath deposition. Prior to buffer deposition, the absorber layers are etched in 5 wt. % potassium cyanide (KCN) for 30 seconds.

Photoluminescence (PL) spectroscopy is carried out in a self-built setup on absorber/buffer stacks, i.e. without deposition of a window layer and thus without the introduction of a space charge region. A 660 nm diode laser is used as excitation source. The PL flux is collected by two off-axis parabolic mirrors and directed via an optical fiber into a monochromator. A grating is used to disperse the PL light onto a 512 element InGaAs detector array. The excitation flus as well as the PL flux is calibrated to absolute numbers as described elsewhere [18]. Some of the PL spectra are influenced by interference fringes. In that case, the PL peak position is estimated using a Gaussian fit, where the fitting range is outside the interference fringes. Supplementary Figure 1 shows the procedure in the presence (a) or absence (b) of interference fringes in the PL spectrum.

The optical diode factor is measured on some samples using excitation dependent PL using Eqn. (1) [19,20]

$$A = \frac{\partial \ln \Phi_{PL}}{\partial \ln G} \tag{1}$$

where $\Phi_{PL}$ is the spectrally integrated PL flux and $G$ is the generation flux. As for the PL measurements, these measurements are carried out on absorber/buffer stacks. As no space charge region is present, the optical diode factor corresponds to the diode factor of the quasi-neutral region.

Time-resolved PL is recorded with a time-correlated single photon counting unit. The sample is excited with a pulsed laser at 10 MHz repetition rate and a wavelength of 640 nm.

Solar cells are finished from the absorber/buffer stacks by sputtering a transparent conductive oxide consisting of a double layer of i:ZnO/Al:ZnO. Subsequently, a Ni/Al grid is deposited by e-beam evaporation.

For each sample, six solar cells are defined by mechanical scribing with areas around 0.5 cm$^2$. Exact areas are determined using a microscope coupled to a digital camera. Parameters of the individual solar cells in this paper are shown only for the solar cells, which have an efficiency of at least 70% of the efficiency of the best solar cell on the same substrate.

Current density voltage (JV) characteristics are measured under a AAA solar simulator at standard test conditions. JV data is fitted following a 1-diode model (Eqn. (2)), where the orthogonal distances are minimized on a logarithmic ordinate axis.

$$J(V) = J_0 \left[\exp\left(\frac{q(V - r_s J)}{A_{el} k_B T}\right) - 1\right] + \frac{V - r_s J}{R_{sh}} + J_{ph} \tag{2}$$

In (2), $J$ is the current density, $V$ the applied bias voltage, $J_0$ is the saturation current density, $A_{el}$ the electrical diode factor, $r_s$ the series resistance, $R_{sh}$ the shunt resistance, $J_{ph}$ the photo-current, $k_B$ the Boltzmann constant, $q$ the elemental charge and $T$ the temperature.

The external quantum efficiency (EQE) is measured in a self-built setup with a lock-in amplifier. No bias voltage or bias light is applied during the measurements. Calibrated Si and InGaAs photodiodes are used as a reference. The photovoltaic bandgap $E_{G,PV}$ is determined from the $dEQE/dE$ peak as suggested by Rau et al. [21]. However, for a number of samples in this study, the EQE is influenced by interference fringes near the absorption edge, which prevents a straight forward application of the procedure to obtain $E_{G,PV}$.Therefore, similar as for the determination of the PL peak position, the $dEQE/dE$ spectra are fitted with a Gaussian curve and the maximum is used for $E_{G,PV}$. Supplementary Figure 1 shows examples of the determination of the $dEQE/dE$ peak position in the absence (a) or presence (b) of interference fringes.

Capacitance-voltage (CV) measurements are carried out in the dark at 300 K, using an ac-modulation voltage of 30 mV. The sample is kept in the dark over night at 300 K so that a relaxed state is obtained.

Transmittance *T* and reflectance *R* spectra are acquired in a photospectrometer using an integrating sphere. The absorption coefficient $\alpha$ is calculated taking into account reflections from a freestanding film [22-24], from which the direct bandgap $E_{g,dir}$ is determined from a Tauc plot following Eqn. (3) [25].

$$\alpha E \propto \sqrt{E - E_{g,dir}} \tag{3}$$

where $E$ is the photon energy.

## 3. Results

Figure 2 shows XRD diffractograms of the (1 1 2) reflection of various $CuIn(Se_{1-x}Te_x)_2$ absorber layers with different compositions *x*. With increasing Te content, the (1 1 2) reflection shifts towards smaller 2Θ values. This is expected due to the larger atomic size of Te compared to Se and thus increased lattice constants [15]. Also, for the complete investigated compositional space it is possible to achieve single phase material as evidenced by a rather narrow (1 1 2) reflections. Elemental depth profiles are obtained from SIMS data. Supplementary Figure 3 shows the profiles for Cu, In, Se, and Te, which are fairly flat throughout the depth of the $CuIn(Se_{1-x}Te_x)_2$ film in agreement with the narrow (1 1 2) reflections. These results show that $CuIn(Se_{1-x}Te_x)_2$ is completely miscible at least up to x = 0.5, which is the relevant compositional range for applications as a bottom-cell. In addition, , it is possible to grow single-phase $CuIn(Se_{1-x}Te_x)_2$ up to at least x = 0.5 using a single-stage process, unlike reported previously where x = 0.2 was given as a limit [16].

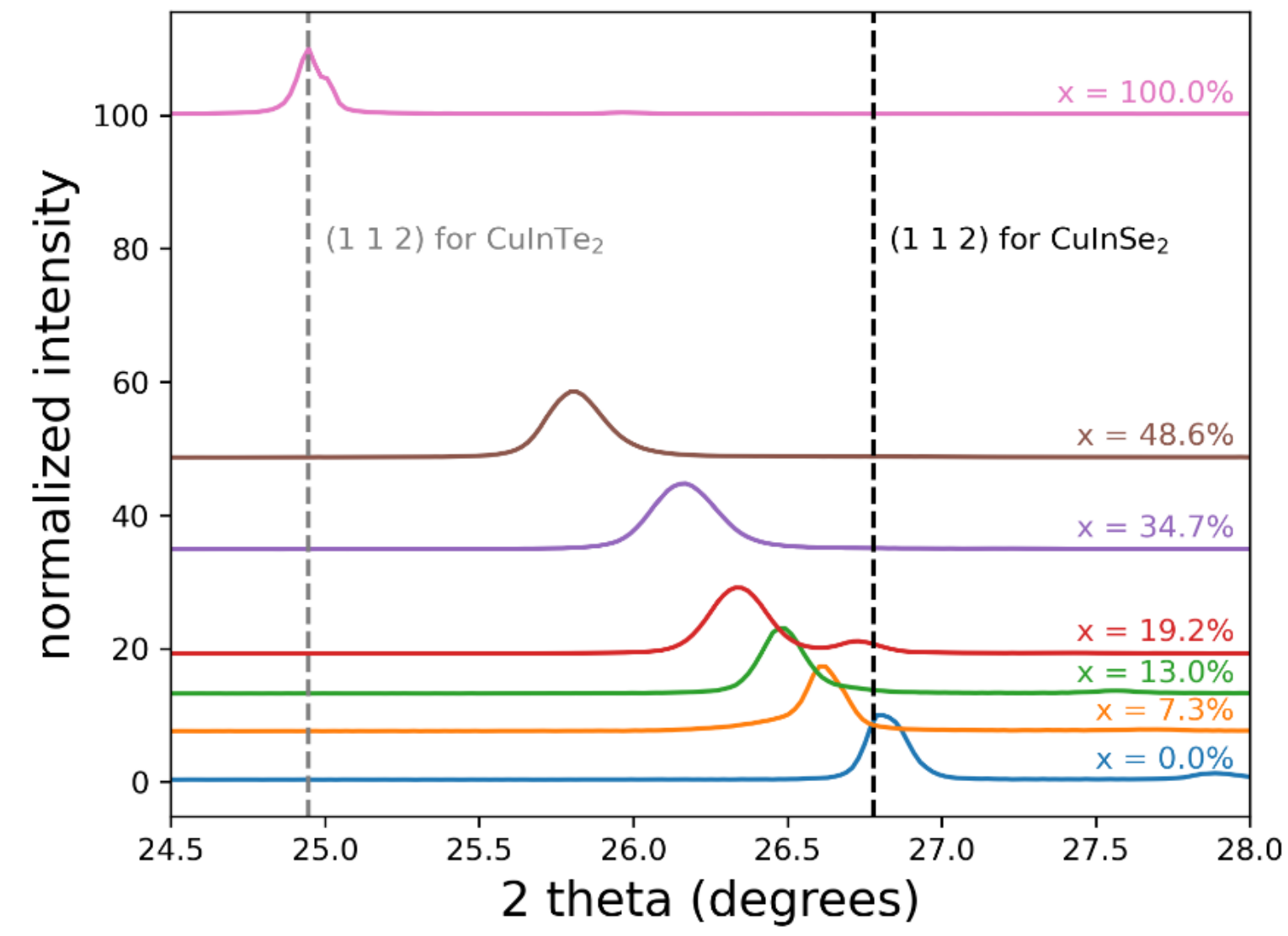

Figure 2 – XRD diffractograms of the (1 1 2) reflection of $CuIn(Se_{1-x}Te_x)_2$ absorber layers – shifted vertically in proportion to the Te content for clarity. Vertical black and gray line indicates the position of the (1 1 2) reflection for the $CuInSe_2$ and $CuInTe_2$ film, respectively.

Photoluminescence spectroscopy is carried out to investigate the properties of absorber layers before forming the pn-junction. Only a CdS buffer layer is deposited on top of the absorber layer for surface passivation [26,27]. The PL peak position reflects the energy difference for the band-to-band transition and thus is an indicator of the bandgap of the absorber layer. In addition, the quasi Fermi level splitting (QFLS) can be determined using Planck's generalized law from the high energy wing of the PL peak [28,29], which represents an upper limit for the open circuit voltage $V_{OC}$. Figure 3 shows PL spectra of selected absorber layers with different compositional values x. Clearly, the PL spectra shift towards lower energies with increasing Te content in the absorber layer in the considered composition range. This red-shift can be linked to a lowering of the bandgap, as expected upon alloying $CuInSe_2$ with Te [9,10]. Another observation is that the PL spectra exhibit multiple peaks, which is strongest pronounced for the sample with x = 7.3 % in Figure 3. These multiple peaks are not necessarily different phases or defect related transitions, but can be caused by interference effects [30]. In order to check if interference effects are the root of these multiple peaks, reflectance spectra are measured on the exact same samples. The PL and the reflectance spectra are both plotted in Supplementary Figure 1. It is shown that the PL peak positions are in good agreement with the peak positions of the reflectance. To support the hypothesis of interference effects, PL spectra are acquired with the sample tilted with respect to the laser beam and thus the direction of PL collection (thus not perpendicular). For such a modification of the sample orientation, the effective optical path of the PL light increases. Therefore, the interference fringes are expected to change as a function of the sample tilt angle [31]. Supplementary Figure 2 shows the PL spectra at two different tilt angles for a sample, which is strongly influenced by interference fringes in standard measurement configuration (excitation and collection perpendicular to the absorber's surface). It is shown that the PL peaks shift when tilting the sample, which is a strong support for the hypothesis that the different PL peaks are caused by interference fringes. Thus, in order to estimate the PL energy for the band-to-band transition, i.e. the peak position, which would be obtained if interference fringes were absent, a Gaussian fit is carried out as detailed in section **Error! Reference source not found.** and exemplarily shown in Supplementary Figure 1. For a l arger set of samples, the PL peak positions from such a Gaussian fit are plotted versus the Te content x in Figure 4 as blue circles. The PL peak position decreases up to $x \approx 0.5$. For $x = 1$, i.e. $CuInTe_2$, the PL peak position is similar to the one of $CuInSe_2$. The reason is the strong bowing for the bandgap of $CuIn(Se_{1-x}Te_x)_2$ [8].

The bandgap is also determined from transmittance and reflectance data obtained by spectrophotometry for samples grown on SLG substrates without a Mo back contact, i.e. on transparent substrates. The extracted bandgap values are shown in Figure 4 as black squares. As for the PL peak position, a decreasing bandgap with increasing Te content up to $x \approx 0.5$ is observed. However, the bandgap values determined from spectrophotometry are roughly 50 meV higher than the PL peak positions. The reason for the red-shifted PL peak positions from their bandgap values can be explained by tail states [32,33]. In particular, it is shown in [32] that the emitted photon spectrum is red-shifted compared to the absorptance spectrum, which explains the 50 meV difference observed here.

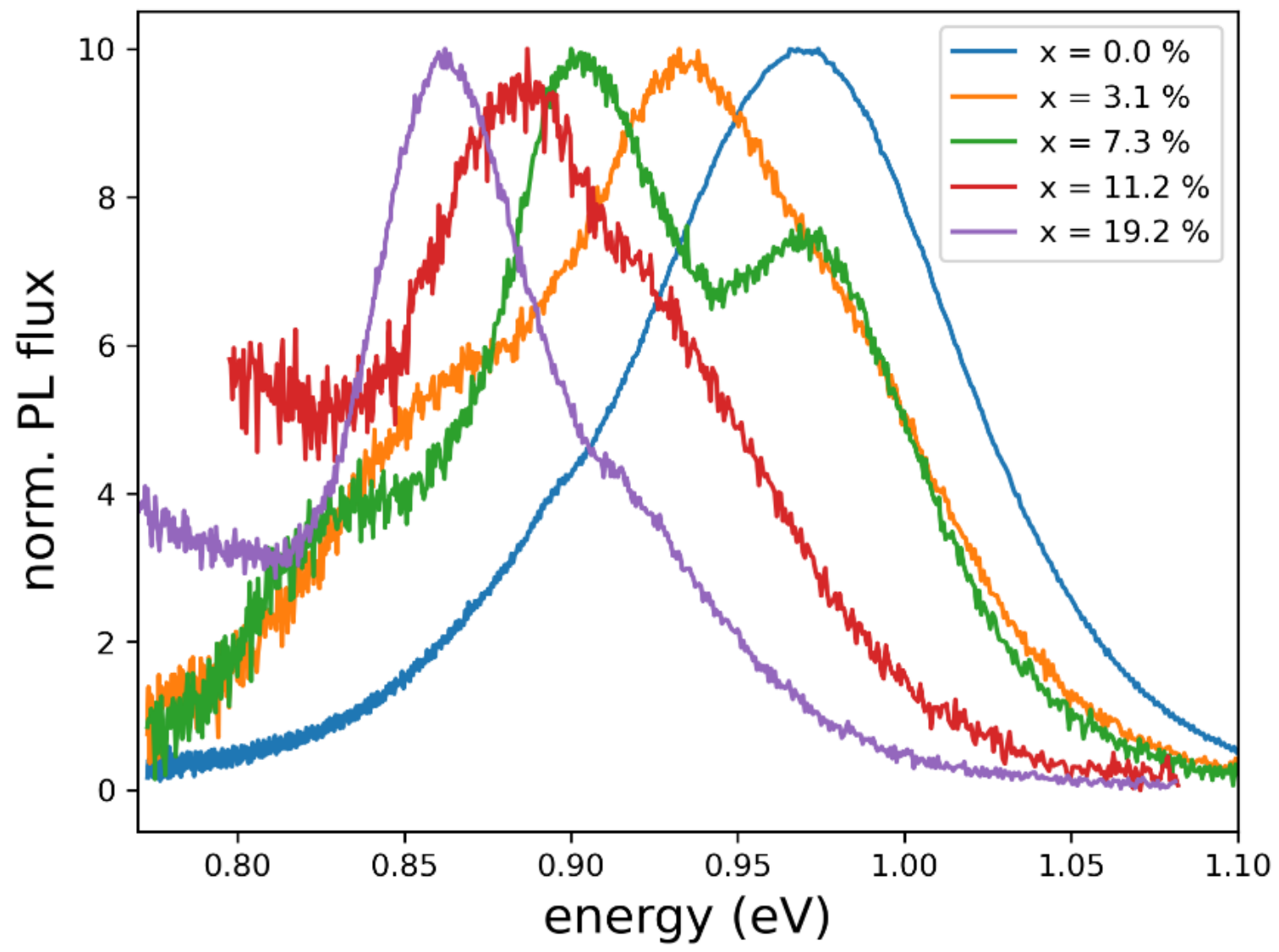

Figure 3 –PL spectra of selected $CuIn(Se_{1-x},Te_x)_2$ absorber layers with different composition *x*.

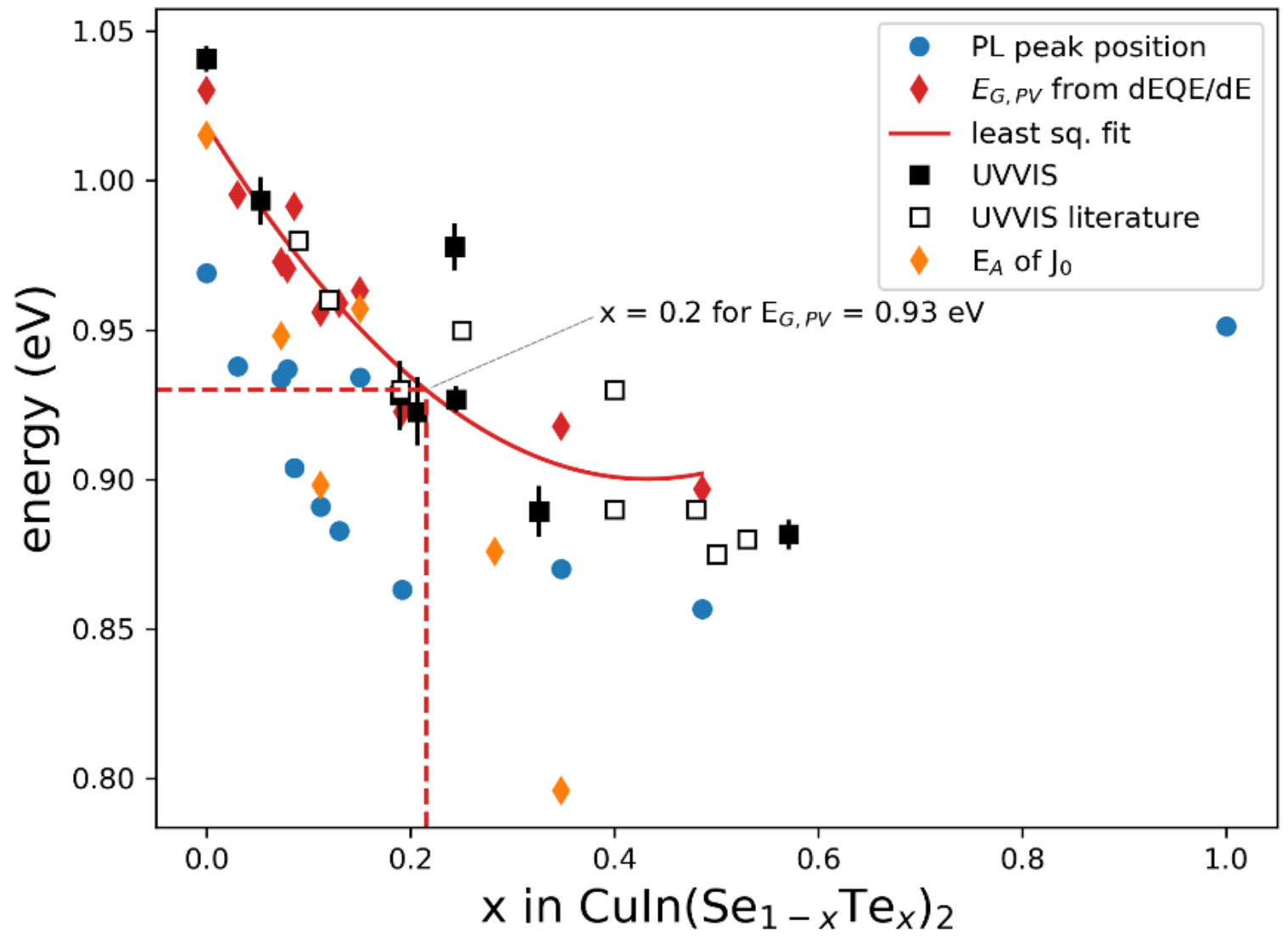

Figure 4 – Estimations of the bandgap energy as a function of the Te content *x*.

Bandgap values $E_{G,PV}$ of the photovoltaic devices are determined by EQE measurements as described in section **Error! Reference source not found.** and are plotted as red diamonds in Figure 4. These bandgap v alues are in good agreement with the values determined by spectrophotometry. This behavior is expected in the presence of bandgap fluctuations, as the EQE is determined by the absorptance spectrum of the thin film solar cell. Bandgap values from literature [10,16] (determined from transmittance and reflectance measurements and the Tauc plot) are plotted as open squares and agree with the values reported here.

JV characteristics are measured for all solar cell devices with different compositions x. The $V_{OC}$'s are plotted versus x in Figure 5 as blue circles. Plots of the efficiency, $J_{SC}$, fill factor, and extracted diode factor are shown in Supplementary Figure 5. The highest $V_{OC}$ of 462 mV is obtained for a $CuInSe_2$ device grown in the same deposition machine. Using the fitted values of $E_{G,PV}$ (red dashed line in Figure 4), expected $V_{OC}$ values are shown in Figure 5 as a red dashed line under the assumption of the same luminescence quantum efficiency (PLQY) in the $CuIn(Se_{1-x}Te_x)_2$ devices as in our $CuInSe_2$ device. Clearly, $V_{OC}$ drops below the expectation as soon as Te is alloyed to $CuInSe_2$. Upon incorporation of Te, $V_{OC}$ drops below 400 mV. For $x \approx 0.5$, $V_{OC}$ drops below 200 mV. The black dashed line is a polynomial fit of order two to the best $V_{OC}$ values of all samples alloyed with Te. Thus, the red area indicates additional non-radiative losses in the solar cells when Te is incorporated.

In comparison, QFLS obtained from absolute PL measurements follow roughly the trend of $V_{OC}$ up to $x \approx 0.25$ (orange diamonds in Figure 5).The QFLS losses (Eqn. (4)) and $V_{OC}$ losses (Eqn. (5)) allow a closer inspection of non-radiative losses as a function of Te content in the absorber.

$$QFLS \text{ loss} = E_g - QFLS \tag{4}$$

$$V_{OC} \text{ loss} = E_g - V_{OC} \tag{5}$$

Smaller values of these loss quantities indicate a lower fraction of non-radiative recombination and thus a superior material quality. Figure 6 shows the QFLS and $V_{OC}$ losses using $E_{G,PV}$ or the PL peak position as the bandgap value $E_g$ in equations (4) and (5). The QFLS losses (Figure 6a) scatter by up to 100 meV, however no clear trend with the Te content x is observed. In particular, various samples exist with similar or even lower QFLS losses compared to the reference sample with x=0. This indicates that the quality of the absorber bulk does not deteriorate and can even improve with the incorporation of Te [1].

In contrast to the QFLS losses, $V_{OC}$ losses (Figure 6b) have the lowest value for $x = 0$, i.e. for $CuInSe_2$. Therefore, additional non-radiative losses are introduced upon finishing the absorber to a solar cell. The increasing difference between QFLS and $V_{OC}$ indicates the occurrence of interface recombination [34]. A few samples are measured using temperature dependent IV (IVT) to determine the activation energy of $J_0$ from extrapolation of $V_{OC}(T)$ to $T=0$ (orange symbols in Figure 4). For some samples, $E_A$ matches well the respective bandgap value $E_{G,PV}$. For two other samples $E_A$ is slightly reduced compared to $E_{G,PV}$ with $E_{G,PV} - E_A \approx 50$ meV. However, the sample with the highest Te content measured by IVT with x = 34.7 % shows a significantly reduced $E_A$. Supplementary Figure 6 shows a correlation of $QFLS - V_{OC}$ versus $E_{G,PV} - E_A$ and supports the claim that additional non-radiative recombination channels at the $CuIn(Se_{1-x}Te_x)_2$/CdS interface are introduced upon finishing the absorber to a solar cell.

[1] Looking at Figure 5 at the red dashed line and the values of the QFLS, it might be surprising that some samples show a smaller QFLS loss (Figure 6a) than the $CuInSe_2$ sample. However, for the data points in Figure 6b the actual measurements of $E_{G,PV}$ (red diamonds in Figure 4) are used, not the interpolated values .

It is interesting to note that a solar cell with $x = 0.4$ and $V_{OC} = 0.293$ V is presented in ref. [16]. In particular, the $V_{OC}$ is ≈ 80 mV higher than the fit of our best $V_{OC}$ values (black dashed line) and thus has a relatively small $V_{OC}$ loss considering a Te content of x = 0.4 (Figure 6b). The $V_{OC}$ value is however similar to the QFLS values we obtain in this compositional range. The device structure for the devices reported in [16] is the same as used in this study. However, a distinct difference in the growth process of the absorber layer is the utilization of a 3-stage process [17] in ref. [16]. How a 3-stage process influences the interface properties and thus avoids the addition of recombination channels upon finishing the solar cell is unclear at the moment and remains to be investigated.

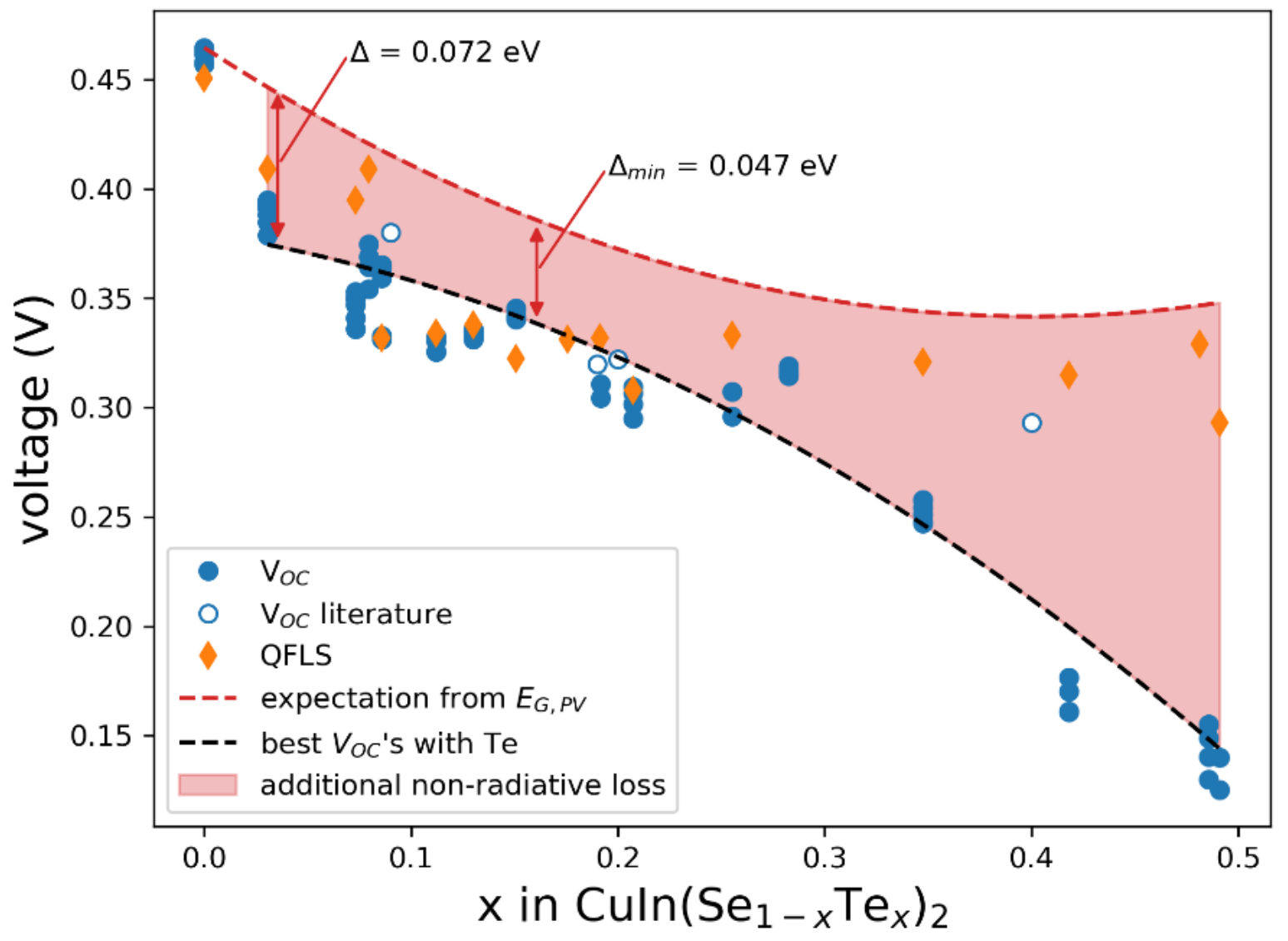

Figure 5 – Open circuit voltages (blue circles) and quasi Fermi level splittings for $CuIn(Se_{1-x}Te_x)_2$ devices and absorber layers. Red dashed curve shows the expected decrease of the $V_{OC}$ upon Te incorporation into $CuInSe_2$ based on the fitted bandgap values $E_{G,PV}$ (see Figure 4), assuming the same PLQY. Black dashed line is a least square fit of a polynomial with order 2 to the best $V_{OC}$ values for each sample, which serves as a guide to the eye for $V_{OC}$ as a function of $x$. Literature data for $V_{OC}(x)$ is taken from [10,16].

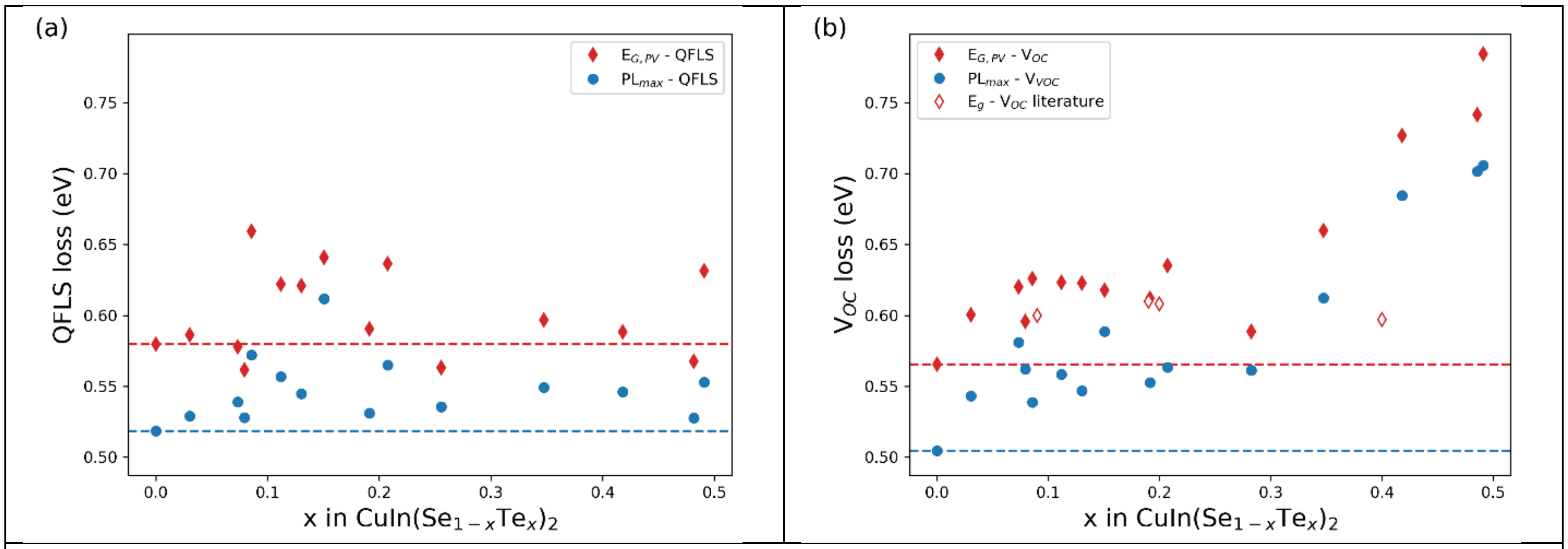

Figure 6 – QFLS losses (a) and $V_{OC}$ losses (b) for $CuIn(Se_{1-x}Te_x)_2$ absorber layers and solar cells as a function of composition x. Horizontal dashed lines indicate the respective loss-value for $x = 0$. Literature data is taken from refs [10,16], where the bandgap is determined from spectrophotometry and a Tauc plot.

Representative JV characteristics and EQE spectra of the best solar cells are shown in Figure 7. The extracted parameters for the performance of the JV curves are listed in Table 1. As already shown in Figure 5, $V_{OC}$ drops upon Te alloying. EQE measurements show a response towards larger wavelengths due to the lowered bandgaps for these solar cells. However, the photo-current is not enhanced by the wider spectral range. Losses in the EQE are caused by incomplete collection as seen by the drop in EQE for wavelengths $\gtrsim$ 900 nm. Losses for lower wavelengths are probably caused by additional recombination channels for Te containing solar cells in agreement with the larger $V_{OC}$ loss (Figure 6b). It is noted that the baseline sample with $x = 0$ does not have an ARC. Hence, losses in EQE are dominated by reflection losses, which is not the case for the two samples with an ARC with $x = 7.9$ % and $x = 15.1$ %. The solar cell with x = 0.079 and a band gap of 0.97eV shows an efficiency of 9.0%, the highest efficiency for a $CuIn(Se_{1-x}Te_x)_2$ device so far.

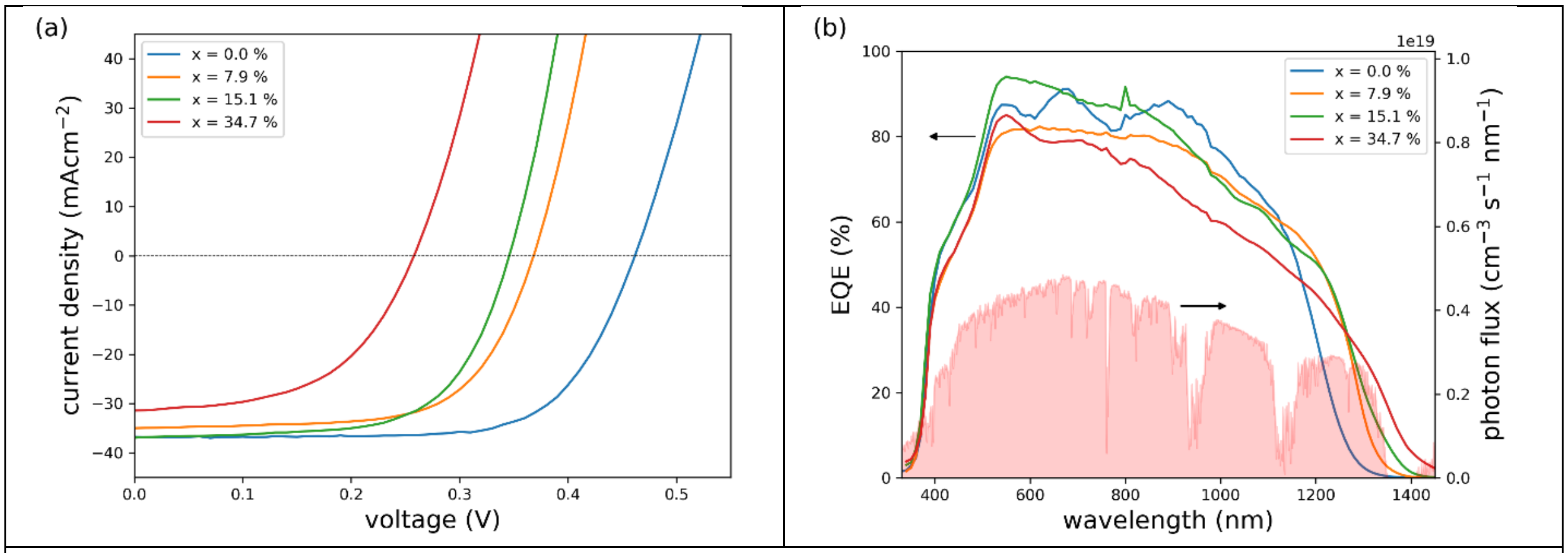

Figure 7 – a) best IV characteristics for the $CuInSe_2$ reference (x = 0) and with Te alloyed solar cells. b) EQE spectra for the same cells as shown in a). Shaded red area on right ordinate shows the photon flux of the AM1.5G spectrum. Note, that the solar cells with x=7.9% and x=15.1% have an anti-reflective coating.

| Table 1 – Parameters of the IV characteristics shown in Figure 7. | | | | | |
|---|---|---|---|---|---|
| | | | | | |
| **composition *x* (%)** | $V_{OC}$ **(mV)** | $J_{SC}$ **(mAcm$^{-2}$)** | $FF$ **(%)** | **Efficiency (%)** | $E_{G,PV}$ **(eV)** |
| 0 | 462 | 36.9 | 69.7 | 11.9 | 1.03 |
| 7.9 [a] | 374 | 36.1 | 66.7 | 9.0 | 0.97 |
| 15.1 [a] | 341 | 37.4 | 64.5 | 8.2 | 0.96 |
| 34.7 | 257 | 31.4 | 53.2 | 4.3 | 0.92 |
| | | | | | |
| [a] with anti-reflective coating | | | | | |

It is encouraging that the QFLS loss is not deteriorating when alloying $CuInSe_2$ with Te (Figure 6a). However, for a doped (p-type for $CuInSe_2$ polycrystalline films) semiconductor, the QFLS can be modified by a change of the doping density or the minority carrier lifetime. The doping density is determined on solar cells from capacitance-voltage measurements as described in the methods section. The doping profiles as a function of voltage are shown in Supplementary Figure 7. The evaluated minimum of the apparent doping density is plotted as a function of the Te content x in Figure 8. The apparent doping density increases by roughly one order of magnitude from $\approx 4 \times 10^{15}$ cm$^{-3}$ for x = 0 to $\approx 5 \times 10^{16}$ cm$^{-3}$ for x = 0.26. Such an increase in the doping density would increase the quasi-Fermi level splitting by approximately 60 mV [35], which however is not observed. A possibility is a deterioration of the minority carrier lifetime such that the quasi Fermi level splitting does not increase with increasing Te content.

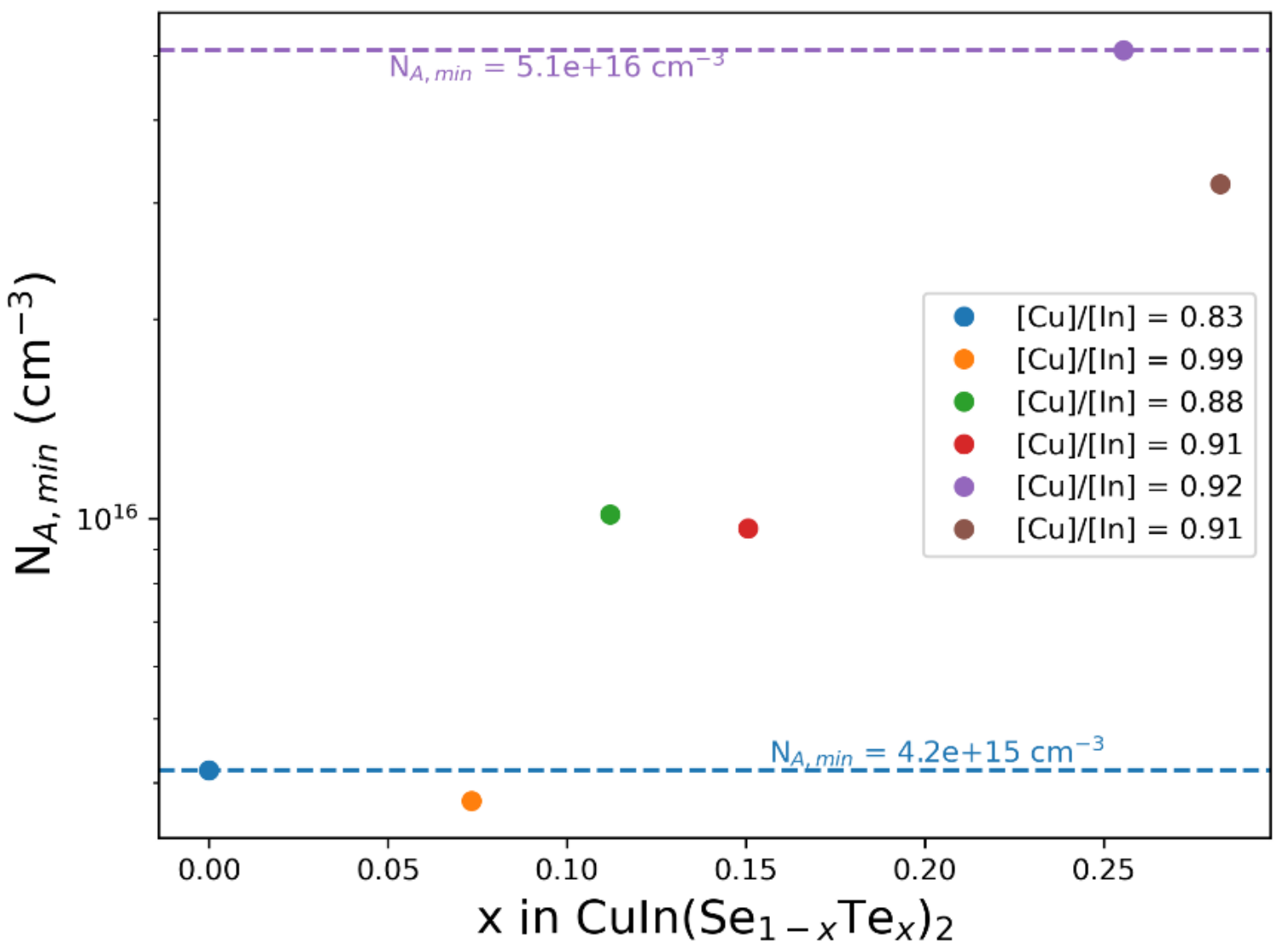

Figure 8 – Apparent doping density obtained from C(V) measurements carried out at 300 K in the relaxed state. Except the sample with almost stochiometric [Cu]/[In] ratio, the apparent doping density seems to increase by almost one order of magnitude when adding Te with $x = 0.25$ to $CuInSe_2$

To estimate the minority carrier lifetime, time resolved PL measurements are carried out. Experimental PL transients for absorbers with various compositions *x* are shown in Figure 9a as semi-transparent lines. The $CuInSe_2$ sample shows a minor initial drop of PL intensity and subsequently follows a single-exponential behavior around 3 ns after the laser pulse. Such a single-exponential behavior is expected for a p-type absorber in low-injection conditions [36]. The initial non-exponential drop might be due to high injection conditions near the absorber's surface just after the injection pulse and/or due to surface recombination. Samples which are alloyed with Te show a much more pronounced initial drop compared to the $CuInSe_2$ absorber and go into a single-exponential behavior roughly 10 ns after the laser pulse. To estimate the tail lifetime (single exponential decay after approximately 10 ns after the pulse) and the magnitude of the initial drop in PL intensity, the transients are fitted with a double exponential function:

$$PL(t) = A_1 \exp\left(-\frac{t}{\tau_1}\right) + A_2 \exp\left(-\frac{t}{\tau_2}\right) \quad (6)$$

In (6), $A_1$ and $A_2$ are the prefactors or amplitudes of the exponential functions and $\tau_1$ and $\tau_2$ the respective lifetimes. Fits to (6) are shown as solid lines in Figure 9a and the fit values $\tau_2$ and $A_2$ for the second exponential are shown in Figure 9b and c, respectively. Only fit values for the second exponential are shown due to several reasons:

i) The second exponential describes the tail lifetime and thus can be identified as the effective lifetime of the minority carriers [37,38]. Note that this lifetime includes effects of bulk and interface recombination.

ii) The initial drop can have various reasons such as: high injection conditions and a subsequent reduction in PL intensity due to reduced bi-molecular recombination and diffusion of charge carriers [39]; minority carrier trapping [40]; space charge regions, which separate charge carriers [41]; enhanced front-surface recombination [39]

Nevertheless, the first exponential is included in the fit (except for the $CuInSe_2$ sample) in order to disentangle the two contributions and to obtain more reliable parameters for the second exponential.

It is observed that the effective lifetime decreases from 22.5 ns for our $CuInSe_2$ sample to approximately 10 – 12 ns for $x > 0$ independent of the Te content (Figure 9b). The amplitude of the second exponential however decreases significantly the higher the Te content in the absorber (Figure 9c). These two phenomena are apparent in Figure 9a as the transients are all more or less parallel for t > 10 ns, but shifted vertically, due to a larger initial drop for higher Te contents. The drop of the effective minority carrier lifetime by a factor of approximately two implies a drop in the QFLS by roughly 20 meV.

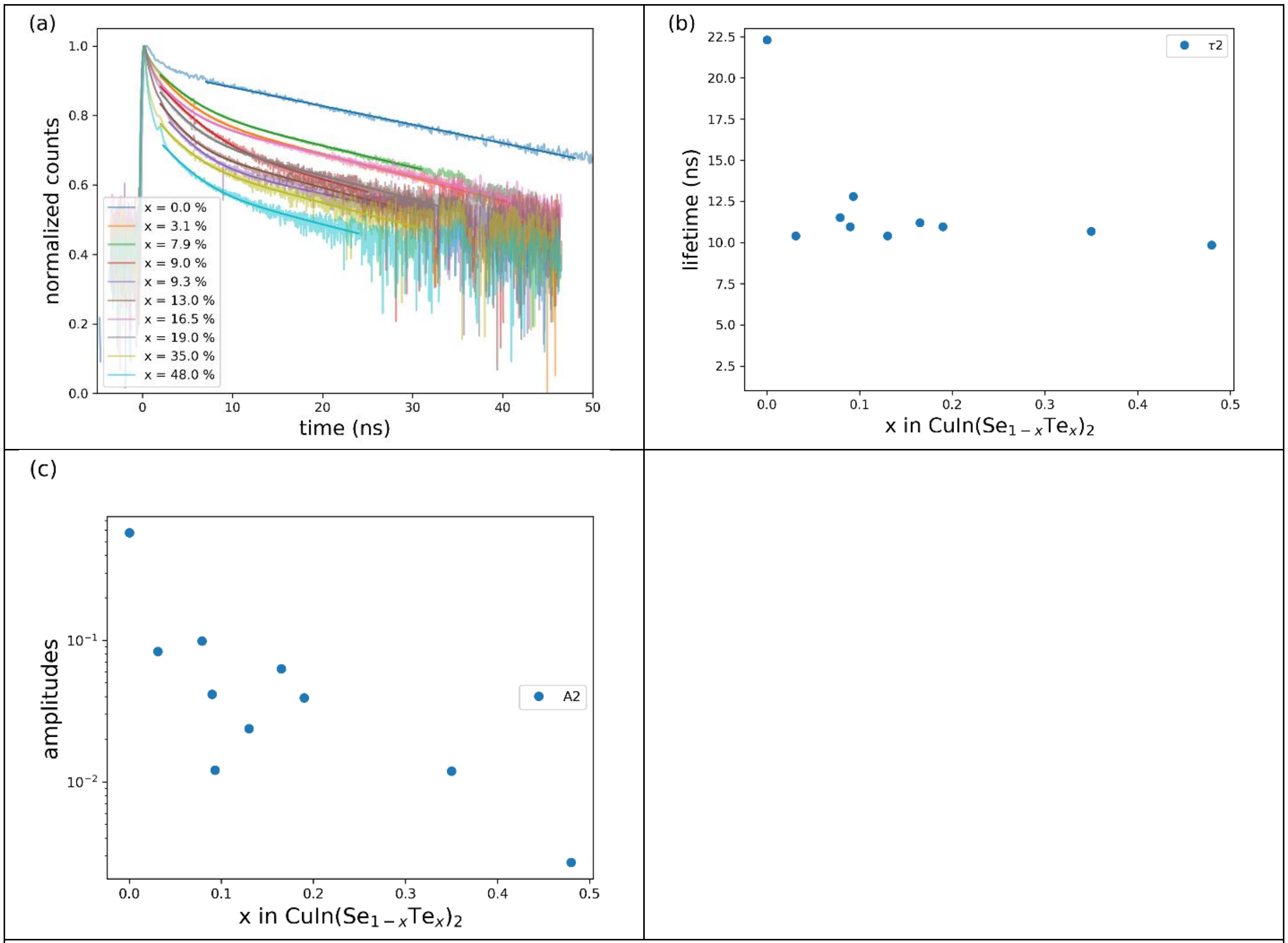

Figure 9 – a) Experimental PL transients for various Te contents x plotted as semi-transparent lines. Double-exponential fits are shown as solid lines. b) Tail lifetime $\tau_2$ of the second exponential as a function of *x*. c) Amplitude of the second exponential as a function of *x*.

# 4. Discussion

In section 3 several losses for $CuIn(Se_{1-x}Te_x)_2$ solar cells are presented and compared to a $CuInSe_2$ reference device. Interestingly, the QFLS is deteriorating very little when introducing Te into the absorber layer as seen in Figure 6a. However, the doping density increases and the minority carrier lifetime decreases with incorporation of Te. The combined effect of these two quantities would imply an increased QFLS by 40 meV, or the QFLS loss would decrease by 40 meV with the addition of Te. However, this is not observed. A possibility might be that the minority carrier lifetime is influenced by trapping [40] or space-charge region effects [41]. Both effects can also explain an increased non-exponential initial drop of the PL signal, i.e. the decrease in the amplitude of the second exponential (Figure 9c). Importantly, the increase of the initial drop is not caused by high-injection conditions, as the doping density increases with Te content (Figure 8), which would rather decrease the range where high injection condition prevail to shorter times. Thus, it is possible that the extracted effective minority carrier lifetimes (Figure 9b) are overestimated (due to trapping or space-charge regions), which might be the reason why the QFLS loss is not increasing with Te content. Additionally, a true minority carrier lifetime smaller than shown in Figure 9b could explain the drop in EQE due to incomplete collection shown in Figure 7b.

Noteworthy, a pure $CuInTe_2$ absorber without Se is also investigated by PL having a QFLS of 456 meV, similar to the value of our $CuInSe_2$ absorber (451 meV). However, neither a PL transient with a response above the instrumental response function (~400 ps) could be measured, nor a functioning device could be fabricated from this absorber layer. We believe that the doping increases significantly for the $CuInTe_2$ absorber, which enables such higher QFLS, while the minority carrier lifetime decreases accordingly. This hypothesis is in agreement with the discussion above.

Next, we focus on the composition of $x = 0.2$ for a bandgap of 0.93 eV. The doping density is around $2 \times 10^{16}$ $cm^{-3}$ based on Figure 8 und thus well suited for a photovoltaic device. It is known that the doping in $Cu(In,Ga)Se_2$ is metastable and that doping densities in the order of $10^{16}$ $cm^{-3}$ are achieved after light soaking [42,43], i.e. in operating conditions. Recently, it was shown that this metastable doping density is responsible for an increased diode factor and therefore a reduced fill factor [18,43]. However, in case of a larger fixed (non-metastable) doping density, a smaller diode factor is expected [43,44]. For the $CuIn(Se_{1-x}Te_x)_2$ solar cells presented here, the relaxed doping density is already in the order of $10^{16}$ $cm^{-3}$ and thus a smaller diode factor due to metastability is expected. Indeed, measurements of the optical diode factor on absorber/buffer stacks, i.e. without a space charge region) yield values between 1.1 and 1.2 (Supplementary Figure 8). These values of the diode factor represent the diode factor in the quasi neutral region. Due to the higher doping density, these diode factors are smaller than the ones generally obtained for $Cu(In,Ga)Se_2$ absorbers around 1.3 [18,20]. This is an important finding because the doping density is less influenced by metastabilities. Thus, solar cells can be realized with low diode factors and therefore relatively large fill factors. For the devices presented here, however, the electrical diode factors (i.e. obtained from the JV characteristics) do not improve significantly (Supplementary Figure 5). Potential reasons might be a decreased inversion in combination with interface recombination, or the addition of SCR recombination.

While the QFLS loss can be maintained, the $V_{OC}$ loss increases with Te content (Figure 6b). IVT measurements and the determination of $E_A$ of $J_0$ indicate that additional non-radiative recombination is introduced at the $CuIn(Se_{1-x}Te_x)_2$/CdS interface. Based on density function theory calculations, Zunger and Wei predicted that the conduction band minimum of $CuInTe_2$ is 0.47 eV higher than the one of $CuInSe_2$ [8]. Thus, it is possible that a cliff-like band-alignment at the absorber/buffer interface is introduced by alloying $CuInSe_2$ with Te. This cliff might be more pronounced for higher Te contents, which is also indicated by the increase in $E_{G,PV} - E_A$ (Supplementary Figure 6). The effect of an enhanced interface recombination, facilitated by the cliff-like band alignment, is always stronger reflected in the $V_{OC}$ loss than in the QFLS loss, because it causes gradients in the minority quasi-Fermi level [34]. In addition, the higher doping density with Te content reduces inversion and may enhance interface recombination [45].

Thus, in order to obtain high quality $CuIn(Se_{1-x}Te_x)_2$/buffer stacks, it is advised to investigate different buffer layers with higher conduction band minima such as (Zn,Sn)O [46,47] or (Zn,Mg)O [48].

In addition, it is worthwhile to study how to control the doping density in these absorber layers. In particular, for 2T tandem applications, a good collection of charge carriers is of importance so that no current-mismatch occurs. Potentially, the amount of Na needs to be better controlled to obtain an optimum in charge carrier collection and $V_{OC}$.

## 5. Conclusion

$CuIn(Se_{1-x}Te_x)2$ absorber layers are fabricated by co-evaporation from elemental sources. Bandgap values as low as 0.9 eV (determined from EQE measurements) are possible for a composition of $x = 0.5$. The QFLS loss, i.e. $E_g - QFLS$, can be maintained upon alloying $CuInSe_2$ with Te, indicating no addition of non-radiative recombination channels. However, the doping density increases, while TRPL measurements indicate a reduction of the effective minority carrier lifetime, i.e. the combined recombination caused by bulk and surface recombination. In contrast to the QFLS loss, the $V_{OC}$ loss increases with the addition of Te, indicating interface recombination. Dominating interface recombination is also indicated by the activation energy of $J_0$ from temperature dependent JV measurements. To avoid the $V_{OC}$ loss with Te incorporation, it is important to optimize the absorber/buffer interface. In particular, it seems that a cliff-like band alignment prevails and is responsible for $V_{OC}$ losses for a sufficiently large Te content in the absorber ($x \gtrsim 0.3$). We believe that alternative buffer layers such as (Zn,Sn)O will enable higher efficiencies for low bandgap absorbers. These absorber layers will then enable an additional degree of freedom for the design of tandem solar cells.

## 6. Acknowledgments

Michele Melchiorre and Thomas Schuler are acknowledged for technical assistance.

# 7. Data availability

The data presented in this paper is available on zenodo.org (detailed doi available before publication)

# Appendix

# A Tandem efficiency calculation

The calculation of the tandem efficiencies shown in Figure 1 are based on the approach described in Ref. [49] for a single-junction solar cell. In particular, the metrics of the current-voltage characteristics are calculated using the following expressions [49]:

$$\Phi_{BB} = \frac{2\pi E^2}{h^3 c^2} \frac{1}{\exp\frac{E}{k_B T} - 1} \tag{A. 1}$$

$$J_{0,SQ} = q \int_{E_g}^{E_{UL}} dE \, \Phi_{BB}(E) \tag{A. 2}$$

$$J_{SC,SQ} = \int_{E_g}^{E_{UL}} dE\Phi_{sun}(E) \tag{A. 3}$$

The upper limit (UL) for the integration is set to $E_{UL}$. For a single-junction solar cell or a top-cell in a tandem configuration, $E_{UL}$ is set to ∞. For a bottom-cell, $E_{UL} = E_{g,top}$, which reflects the fact that no photons from $\Phi_{sun}$ enter the bottom cell after perfect absorption by the top cell.

$$V_{OC,SQ} = k_B T \ln\left(\frac{J_{SC,SQ}}{J_{0,SQ}} + 1\right) \tag{A. 4}$$

The fill factor is determined from the maximum power point (MPP) as

$$FF = \frac{P_{MPP}}{J_{SC,SQ}V_{OC,SQ}} \tag{A. 5}$$

using the following current-voltage characteristics

$$J(V) = J_{0,SQ}\left(\exp\left(\frac{V}{k_B T}\right) - 1\right) - J_{SC,SQ} \tag{A. 6}$$

Finally, the efficiencies for the 2T and 4T configuration are calculated. In the case of the 4T device, the efficiencies for the top- and bottom-cell are added, where the individual efficiencies $\eta$ are obtained via

$$\eta = \frac{V_{OC,SQ}J_{SC,SQ}FF}{P_{sun}} \tag{A. 7}$$

In the case of the 2T device, $J_{SC,SQ,2T}$ and $J_{0,SQ,2T}$ need to be determined first, which is done according to

$$J_{SC,SQ,2T} = \min(J_{SC,top}, J_{SC,bottom}) \tag{A. 8}$$

$$J_{0,SQ,2T} = \frac{J_{0,top}J_{0,bottom}}{J_{SC,SQ,2T}} \tag{A. 9}$$

Expression (A. *8*) enforces *current matching*, i.e. the overall current density is limited by the smaller current density due to the series connection of the cells. Expression (A. *9*) results in a $V_{OC,2T}$ determined by the addition of the $V_{OC}$'s of the individual sub-cells. Subsequently, the $FF$, $J(V)$ and the efficiency is calculated by (A. *5*) and (A. *6*) as described above.

# Supplementary Information

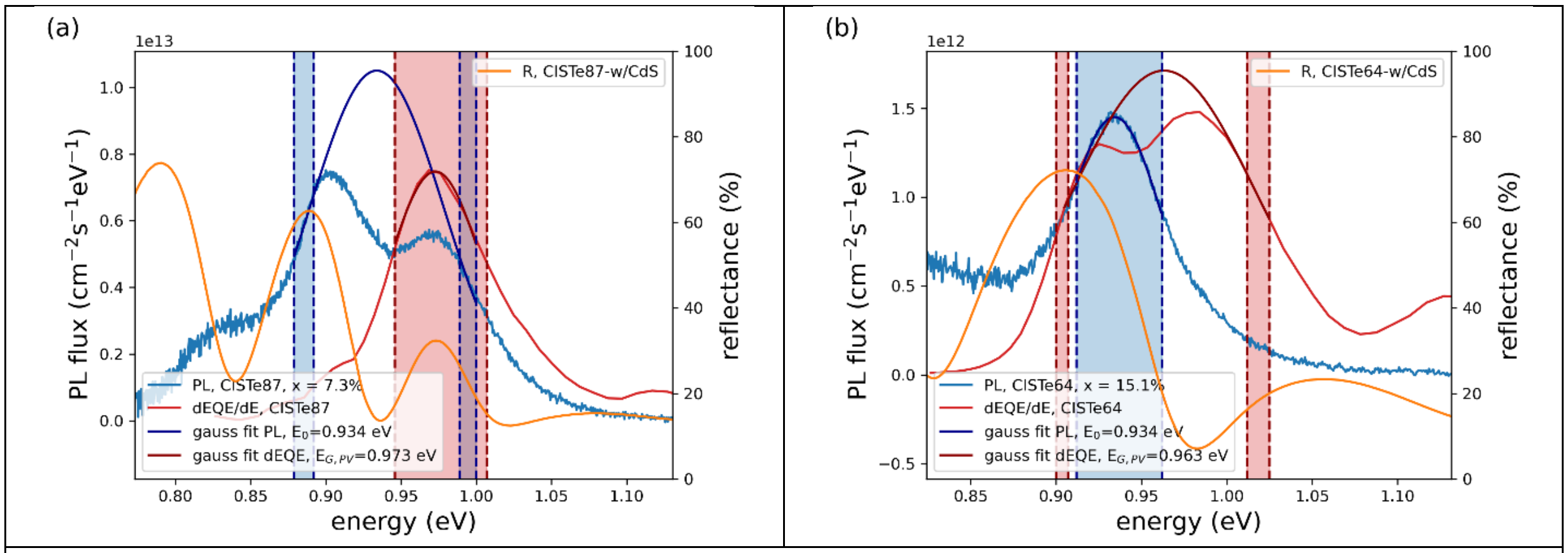

Supplementary Figure 1 – a) PL spectrum of an absorber/buffer stack with x = 7.3 % showing two main peaks (blue curve, left ordinate). The reflectance spectrum (orange curve, right ordinate) is shown as well. A good agreement of the reflectance peaks and the PL peaks is observed, indicating that the double PL peak is indeed caused due to interferences. The PL peak maximum is determined from the center of a gaussian fit (dark blue curve) of the PL data. The fitting range is only outside the double peak at lower and higher energies as indicated by the vertical blue bars. The absorption edge of the $dEQE/dE$ spectrum of the final solar cell device is much less influenced by interference fringes. The maximum of this spectrum is determined from a Gaussian fit where the fitting range (vertical red lines) covers the center of the $dEQE/dE$ peak position. b) PL spectrum in the absence and $dEQE/dE$ spectrum in the presence of interference fringes.

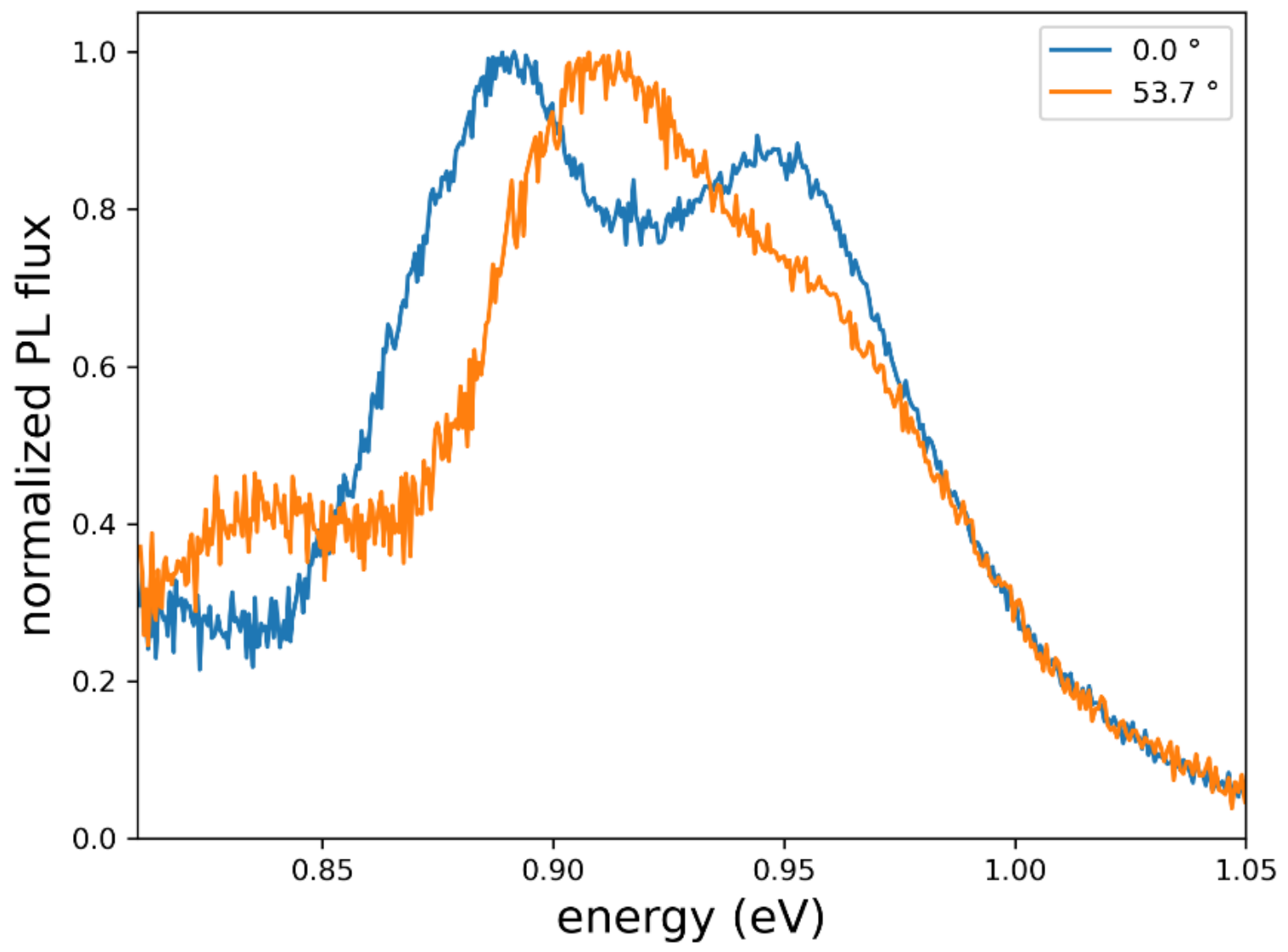

Supplementary Figure 2 – PL spectra for the sample with x = 7.3 % shown in Supplementary Figure 1 acquired under different orientations of the sample with respect to the laser beam and the direction of PL collection [31]. The PL spectra change upon changing the sample orientation, which is an additional indication that the different peaks of the PL spectrum are caused by interference fringes [31].

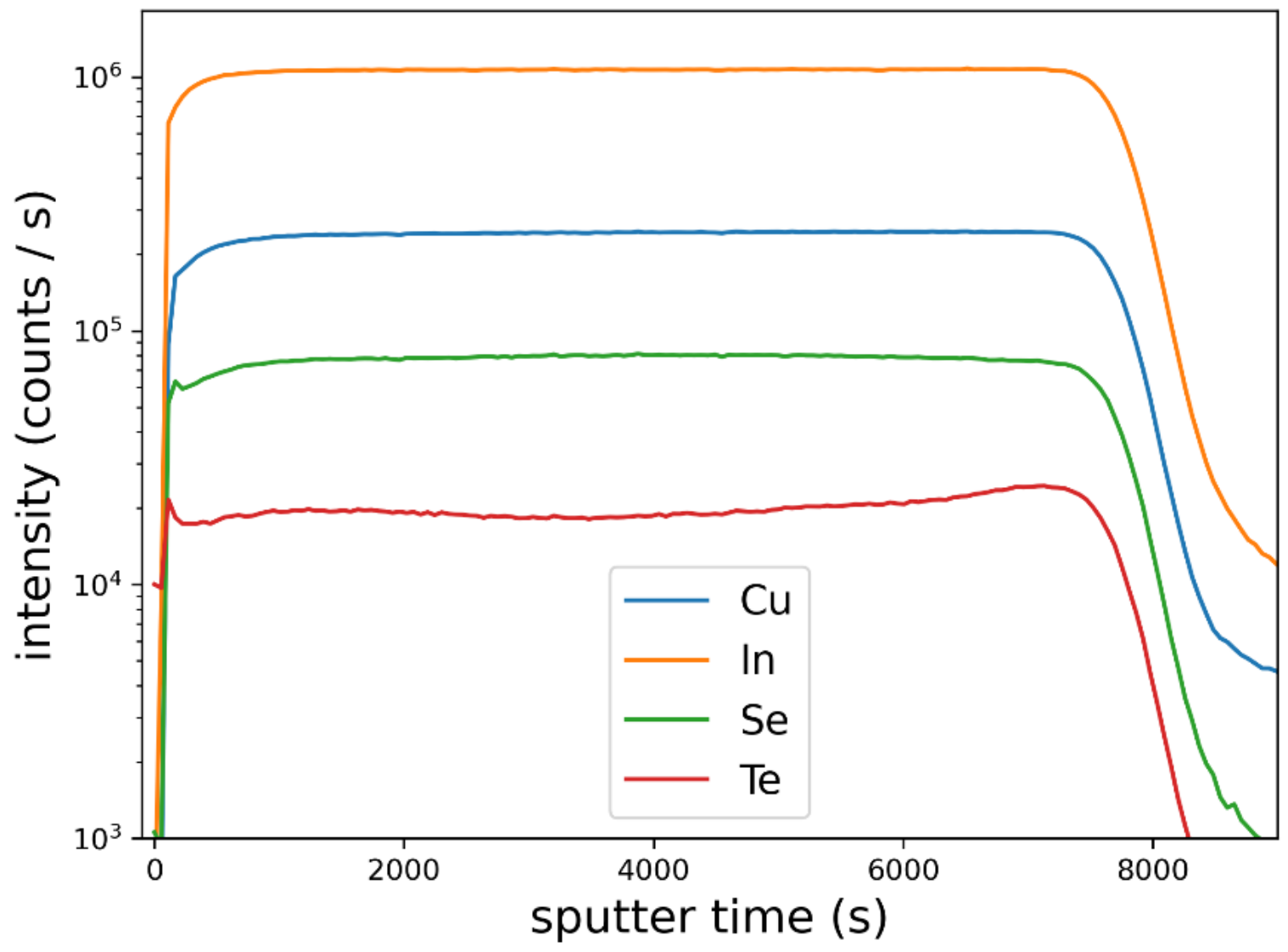

Supplementary Figure 3 – Elemental depth profiles obtained from SIMS data for a sample with x = 15.1 %. The profiles for Cu, In and Se are all very flat. Small variations in the Te profile are visible, which are probably due to a non-constant flux during the deposition of the $CuIn(Se_{1-x}Te_x)_2$ thin film.

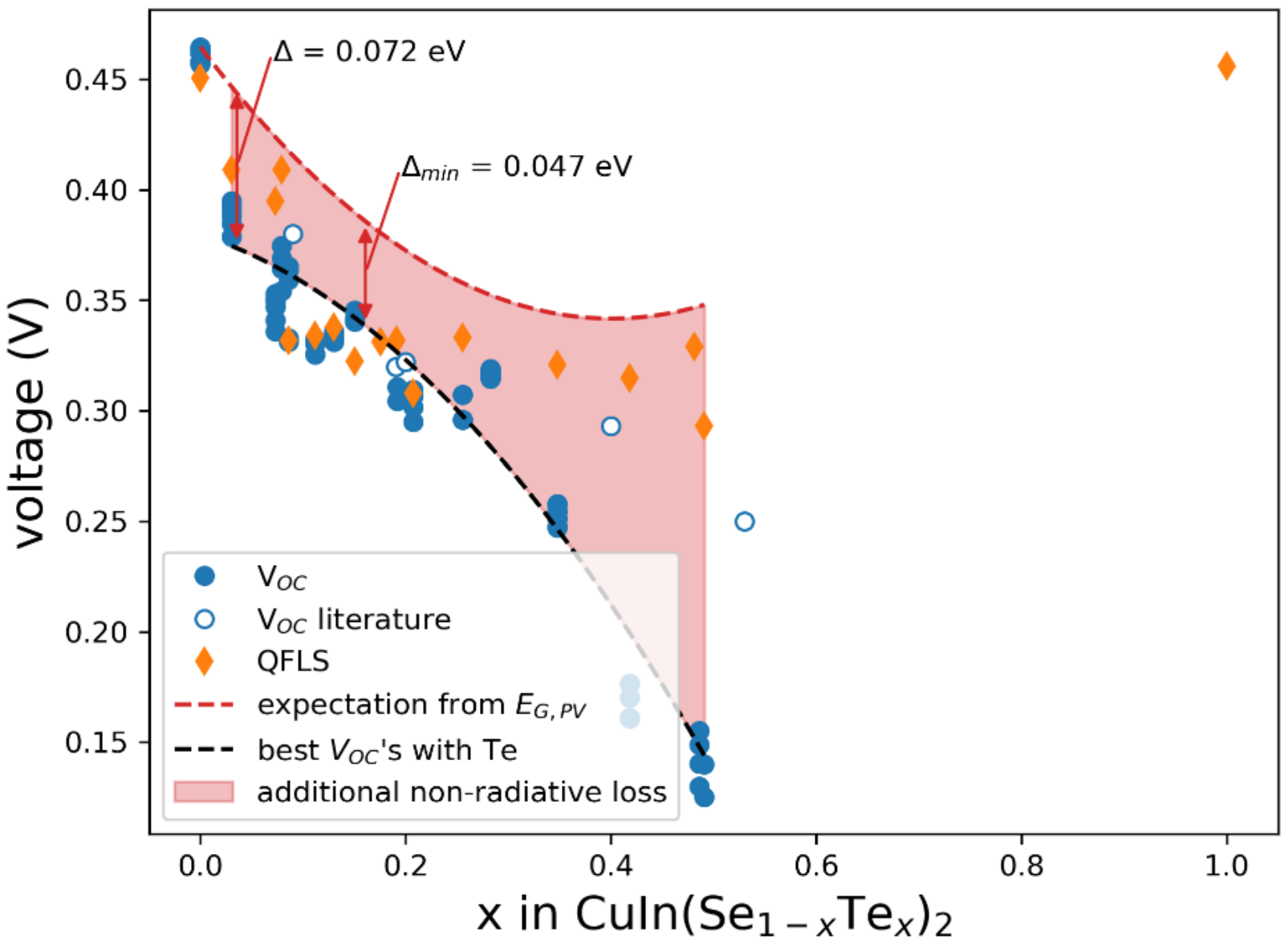

Supplementary Figure 4 – Same data as shown in Figure 4 but over the entire compositional range. Noteworthy is the quasi Fermi level splitting of 456 meV for the $CuInTe_2$ sample ($x = 1$), which is similar to the quasi Fermi level splitting of 451 meV for the $CuInSe_2$ sample ($x = 0$).

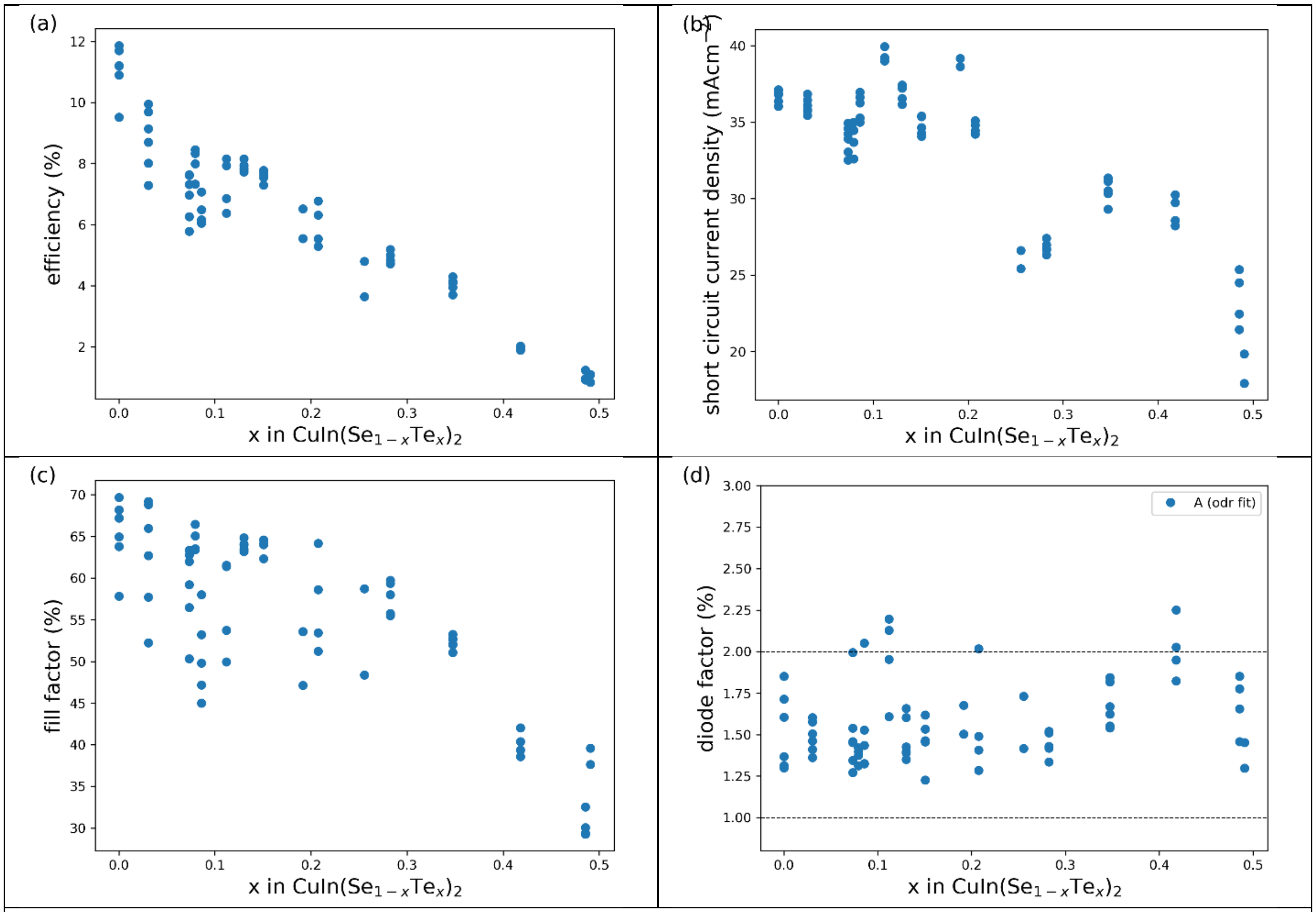

Supplementary Figure 5 – IV parameters as a function of Te content *x* in the absorber layer.

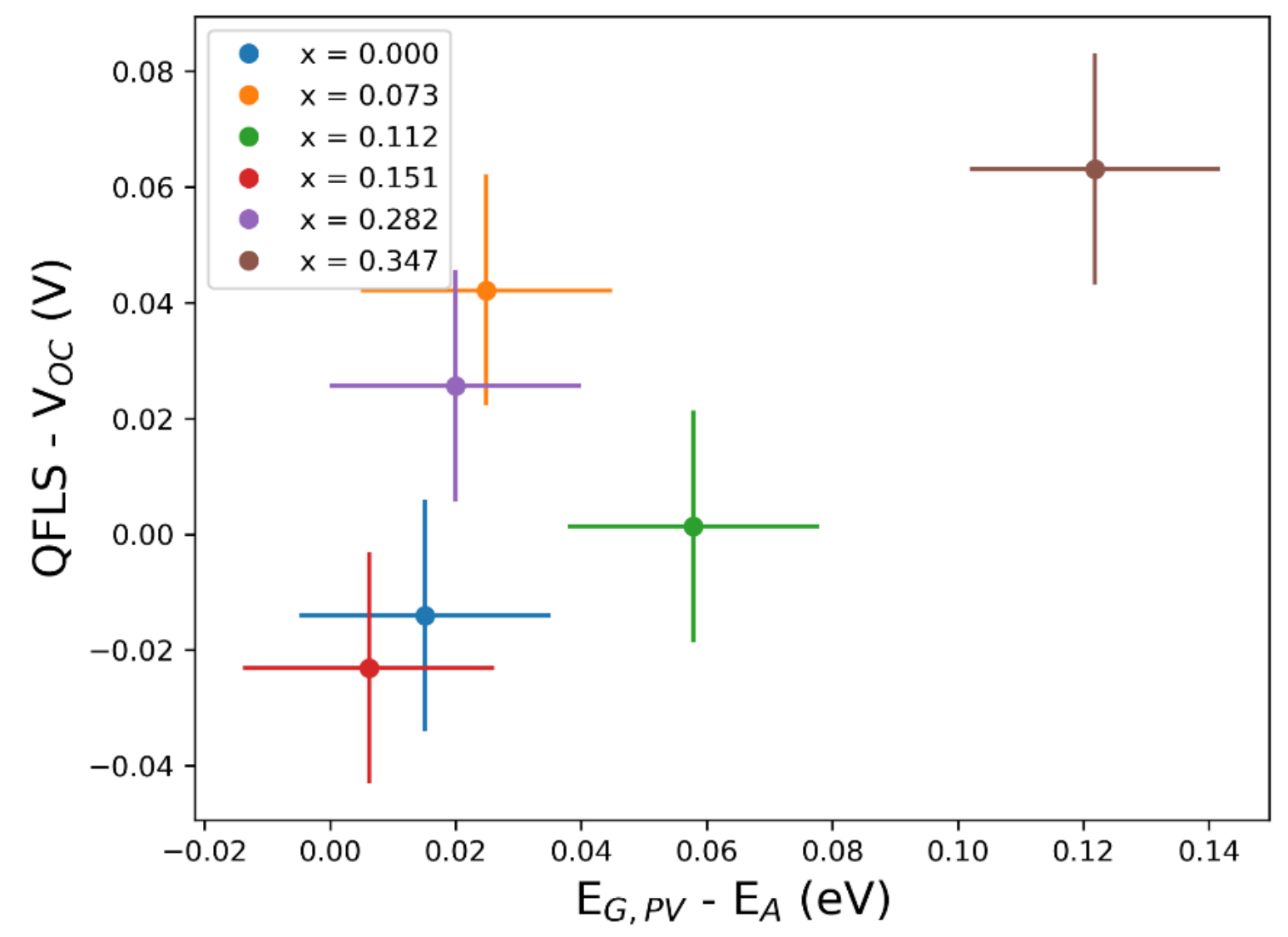

Supplementary Figure 6 – Correlation of the differences between $E_{G,PV} - E_A$ and $QFLS - V_{OC}$. For both quantities an error of 20 meV is assumed.

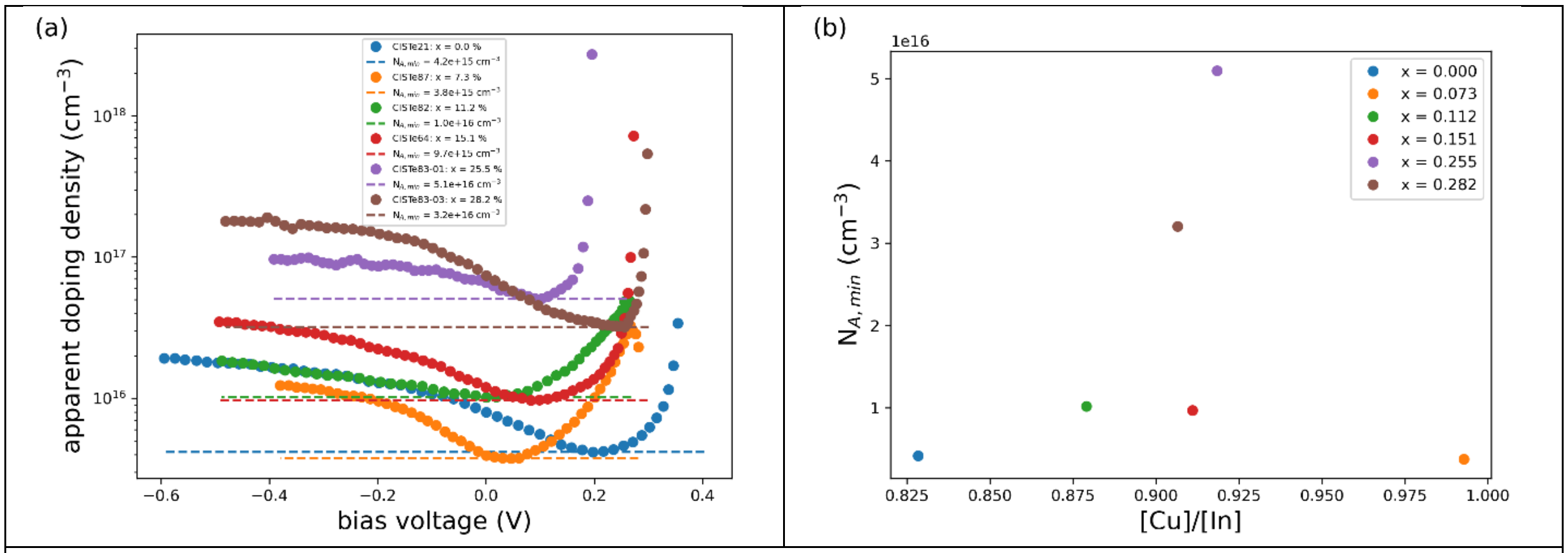

Supplementary Figure 7 – a) apparent doping density as a function of applied bias voltage. Horizontal dashed lines indicate the minimum of the apparent doping density $N_{A,min}$. b) the plot of of $N_{A,min}$ as a function of the [Cu]/[In] ratio shows no clear correlation.

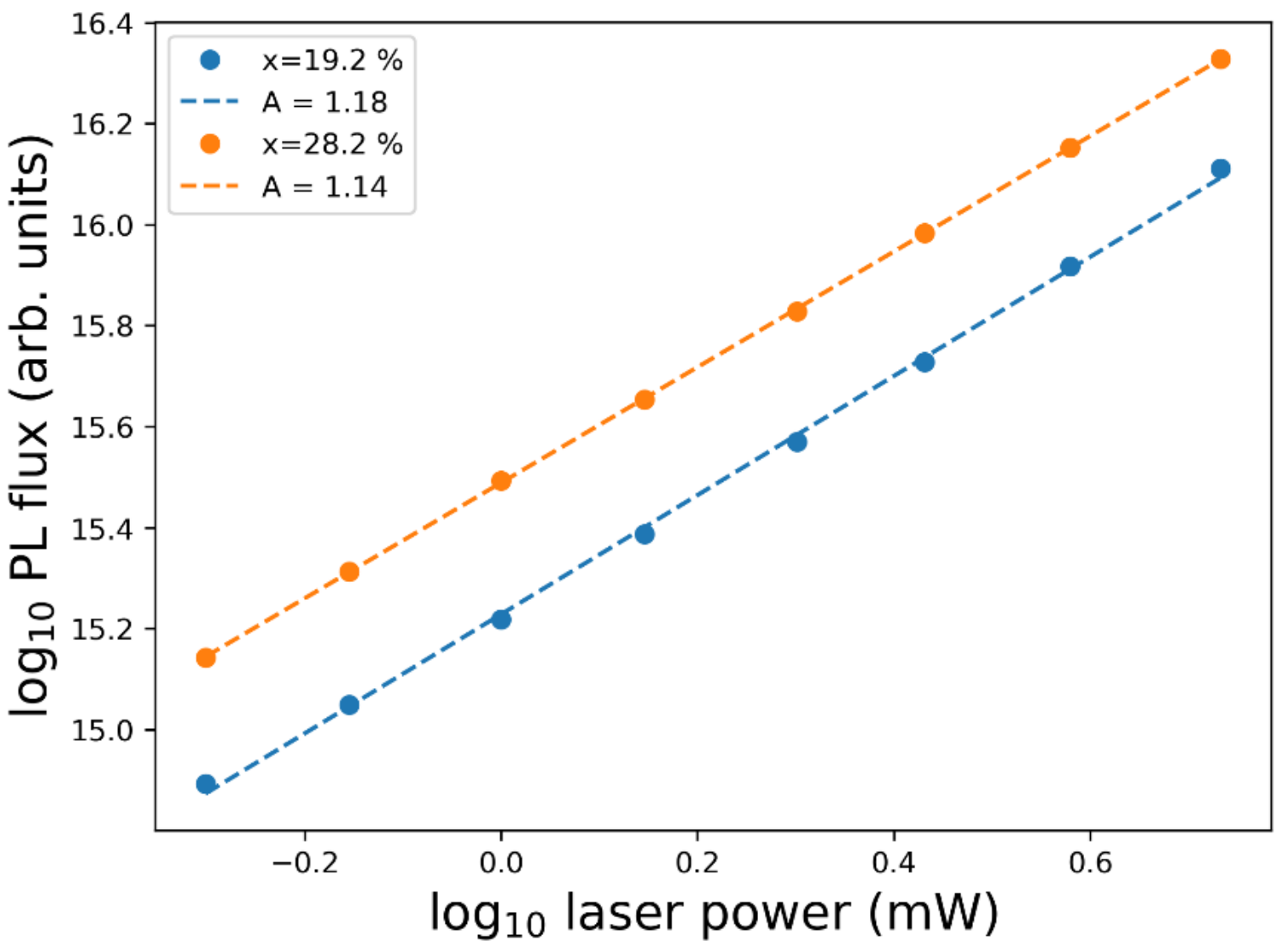

Supplementary Figure 8 – Measurement of the (optical) diode factor $A$ using excitation dependent PL spectroscopy [20] on absorber/CdS layers, i.e. without a space charge region. The slope gives the diode factor of the absorber layer in the quasi neutral region.